\documentstyle[12pt,psfig]{article}

\begin{document}                                                                                   
\title{Gamma Ray and Hadron generated \v Cerenkov Photon Spectra at Various Observation Altitudes} 
\author{M. A. Rahman, P. N. Bhat, B.~S. Acharya, V. R. Chitnis,\\
P.~Majumdar and  P. R.  Vishwanath \\
{\it Tata Institute of Fundamental Research,} \\
{\it Homi Bhabha Road, Colaba, Mumbai 400 005, India.}}
\setlength\textheight{9.0in}
\setlength\topmargin{-1. cm}
\maketitle
\date{}
\baselineskip=15pt
\begin{abstract}

We study the propagation of \v Cerenkov photons generated by Very High
Energy $\gamma -$rays and hadrons in the atmosphere. The photon production
height distributions are estimated from semi-empirical methods and
compared with those derived by standard simulation techniques. Incident
spectra at various observation altitudes are then derived after applying
wavelength dependent corrections due to photon attenuation in the
atmosphere during the propagation of photons from the height of production
to the height of observation. These are generated both for $\gamma -$ and
hadron primaries of various energies. The derived
production height distributions agree very well with those generated by
the simulation package `CORSIKA' at all energies and for both $\gamma
-$ray and proton primaries. The incident photon spectra are found to be
both altitude dependent and primary energy dependent. The peak of
the incident spectrum shifts towards the shorter wavelength with
increasing altitude of observation for a given primary. Also the peak of
the photon spectrum shifts towards the shorter wavelength with
increasing energy of the primary at given altitude. The fraction of the UV
component in the incident \v Cerenkov spectrum is estimated both for
$\gamma -$ray and hadronic primaries at various observation altitudes and
energies. Hadron generated \v Cerenkov spectra are marginally richer in UV
light and the difference increases slightly at higher altitudes. The fraction 
of the UV to the visible light in the \v Cerenkov spectrum could be a useful
parameter to separate $\gamma -$rays from cosmic ray background only if one
can measure this fraction very accurately.

\end{abstract}

\section{Introduction}

In the highest energy end of the electromagnetic spectrum, the energy
range between 50 keV and 30 GeV has been successfully covered by satellite
based missions\cite{har99}. At higher the energies steeply falling spectrum
makes it impossible to observe the $\gamma -$ray sky by satellite based
detectors.
The ground based atmospheric \v Cerenkov technique has proved to be the
most sensitive technique in exploring the celestial $\gamma -$rays in the
energy range 300 GeV - 10 TeV. The basic detection technique in this case
is to deploy one or several parabolic mirrors at the observation level
fitted with fast photo tubes at the focus.  Very high energy celestial
$\gamma -$rays initiate an electromagnetic cascade in the atmosphere as
they enter the atmosphere. The electrons and positrons in the cascade,
being relativistic, emit \v Cerenkov light as they propagate down the
atmosphere resulting in a faint ($\sim 10^{-4}$ times the brightness of
the star light background) flash of light lasting a few nanoseconds. This
fast \v Cerenkov flash is detected electronically by coincidence
technique. However the main drawback of the technique is the presence of
much more abundant cosmic ray background which severely limit the
sensitivity of this technique. Various ingenious techniques are employed
to distinguish between the \v Cerenkov light produced by cosmic $\gamma
-$rays from that by the cosmic rays. Among these, the imaging
technique, wavefront sampling technique, angular and spectral separation
techniques are currently employed by different experiments
\cite{cr93,sc96,fe97,on98}.

In spite of the necessity of carrying out these observations during clear,
moon-less nights which scales to only about 10-15\% duty cycle, the
technique has been successful mainly due to two reasons. Firstly, a large
collection area which is $\sim 10^4 - 10^5 $ times larger than
the satellite detectors provides a high event rate. Secondly, with high
angular resolution and some recently developed $\gamma -$ hadron
separation techniques \cite{fe97,cb00} provide high sensitivity to the
method. Large ground based telescopes based on this technique are now
being built to cover the intermediate energy range of 30 -
300 GeV \cite{tum90,cel96, hof97,wee97}.

In the absence of standard mono-energetic beams of cosmic rays or $\gamma
-$rays, one has to depend on the simulation techniques to understand and
optimize the detector response to these radiations. For this purpose we
have carried out detailed simulations with VHE $\gamma -$rays and cosmic
rays of various primary energies initiating extensive air showers in the
atmosphere. In order to ensure the reliability of the conclusions drawn
from these studies it is imperative to verify the correctness of the basic
results derived from these simulations. There are a few simulation
packages painstakingly developed by high energy physicists and we chose
one of them called CORSIKA \cite{hec98} developed by the KASCADE
collaboration.

CORSIKA (version 562), \cite{hec98,kn98} has been used here to simulate \v
Cerenkov light emission in the earth's atmosphere by the secondaries of
the extensive air showers generated by cosmic ray primaries or
$\gamma-$rays. This program simulates interactions of nuclei, hadrons,
muons, electrons and photons as well as decays of unstable secondaries in
the atmosphere. It uses EGS4 code \cite{ne85} for the electromagnetic
component of the air shower simulation and GHEISHA\cite{fe85} for the
simulation of hadronic interactions at TeV energies. The \v Cerenkov
radiation produced within the specified bandwidth (300-550 $nm$) by the
charged secondaries is propagated to the ground.  The US standard
atmosphere parameterized by Linsley \cite{us62} has been used. The
position, angle, time (with respect to the first interaction) and
production height of each \v Cerenkov photon hitting the detector at the
observation level are recorded.

We have mainly used Pachmarhi (longitude: 78$^{\circ}$ 26$^{\prime}$ E,
latitude: 22$^{\circ}$ 28$^{\prime} N$ and altitude: 1075 $m$) as the
observer's location where an array of 25 \v Cerenkov detectors each of
area 4.35 $m^2$ is deployed in the form of a rectangular array. We have
assumed 17 detectors in the E-W direction with a separation of 25 $m$ and
21 detectors in the N-S direction with a separation of 20 $m$.  This
configuration, similar to the Pachmarhi Array of \v Cerenkov Telescopes
(PACT) \cite{pnb98} but much larger, is chosen so that one can study the
core distance dependence of various observable parameters. Simulations
were also carried out at sea level and a location at an altitude of 2 $km$
above mean sea level. Mono-energetic primaries consisting of $\gamma-$
rays and protons incident vertically on the top of the atmosphere with
their cores at the centre of the array have been simulated in the present
study.

The longitudinal shower development profiles of \v Cerenkov photons in the
atmosphere have been generated using CORSIKA for $\gamma -$ray and proton
primaries of various energies.  Tracks of charged particles are followed
through in steps $20~g~cm^{-2}$ \cite{cb98,cb99}. 

In the present work the \v Cerenkov photon growth curves have been
obtained from the simulations for proton and $\gamma -$ray primaries of
various energies and for the three observation levels mentioned above.
These have been compared with those derived from independent empirical
relations. The \v Cerenkov photon spectra as seen at the observation
levels are then derived for various primary energies of $\gamma -$rays and
protons after taking into account of the wavelength dependent attenuation
in the atmosphere. Then we estimate the relative fraction of UV photons in
these spectra as a function of primary energy and observation level.

\section{Calculation of longitudinal profiles}

\subsection{$\gamma -$ray primaries}

For showers initiated by photons of energy $E_{\gamma}$ the longitudinal
shower development curve is derived from the equation for cascade curve
under approximation A. The average number of electrons
of all energies
$N(E_{\gamma},x)$ as a function of atmospheric depth $x$, is given
approximately, in the region where the number of particles is large, by
\cite{gr56}:

$${{N(E_{\gamma},x)}} = {{0.31} \over \sqrt{\ln \left({E_{\gamma}} \over {E_{t}}\right)}}\exp{(t(1-1.5 {\ln {s}))}}$$
where $t$ is the depth $x$ measured in units of radiation length in the
atmosphere (37.2 $g~cm^{-2}$) and $E_t$ is the the electron critical
energy at the depth $t$. This is the energy below which the electron
multiplication stops.  It is approximately the energy at which radiation
losses and collisional losses are equal and has a value 84.2 MeV in air.
The shower age parameter, $s$, which is a measure of the stage of the
longitudinal shower development, is given by:

$${s} = {{3t}\over {(t+2\ln{E_{\gamma}\over E_t})}}$$
$s$ = 1 at shower maximum while $s<1$ above the shower maximum and and
$s>1$ below. The relation between the atmospheric height $h$, (as measured
from sea level in $m$) and the depth, $x$ ($g~cm^{-2}$ as measured from
the top of atmosphere) is taken from Rao\cite{sr81}:

$${h} = {(6740 + 2.5x)}\ \ln\left( {{{1030}\over {x}}}\right)\indent m$$
where the scale height is dependent on the atmospheric depth. The
refractive index $n$ at a given height $h$ is assumed to be: 

$$n=1+n_0$$
where $n_0$ is given by:

$$n_0 = 2.9\times 10^{-4}e^{-{{h}\over {7100}}}$$

For calculating the electron growth curve the atmosphere is divided
into slabs of thickness 333 $m$, after the height of first interaction of
the primary. The electron threshold energy for the production of \v
Cerenkov radiation at a depth $t$ is given by

$${E_{th}} = \left({{{1}\over {\sqrt{1-(1+n_0)^{-2}}}}-1}\right) m_e$$
where $m_e$ is the electron rest mass (0.51 MeV). The fraction of
electrons with energy above the \v Cerenkov threshold is given by
\cite{ly90}:

$$f_{\check{C}} =\left( {{0.89 E_0-1.2}\over {E_0+E_{th}}}\right) ^s(1+0.0001 s E_{th})^{-2}$$
where $E_0$ = 26 \emph{MeV} when $s\le 0.4$\\
\indent \indent \indent~ = $44-17(s-1.46)^2~~ MeV$ for $s>0.4$\\

The number of electrons above the \v Cerenkov threshold is then computed
in each of the slabs. For each straight section of the electron path, the
number of \v Cerenkov photons produced per unit path-length is given by,

$${{dN}\over {dl}} = 6.28\alpha \left( {{{1}\over {\lambda _1}}-{{1}\over{\lambda _2}}}\right) \left( {1-{{1}\over {\beta ^2n^2}}}\right)$$
in the wavelength band bounded by $\lambda _1$ and $\lambda _2$ and
$\beta$ is the electron velocity and $\alpha$ is the fine structure
constant. Assuming all the electrons in the cascade to be close to the
shower axis and they travel with the velocity close to that of light in vacuum, 
the number of \v Cerenkov photons produced in a path-length $dl$ (= 333 $m$ in
the present case) and in the wavelength range 300-550 $nm$ at a depth
$x~g~cm^{-2}$ is then given by,

$$N_{ph} ={0.402}{N}\left( {{{x}\over {1030}}}\right) dl$$

\subsection{Proton primaries}

Because of the differences in the kinematics of shower development in the
case of hadronic primaries the longitudinal development profiles too are
different as compared to that of $\gamma -$ray primaries. Using the
scaling model for nuclear interactions, Gaisser \& Hillas \cite{gh77} find
that the average number of particles, $N(E_p,x)$ at an atmospheric depth
of $x~g~cm^{-2}$, measured from the first point of interaction, in a
shower initiated by a proton of energy $E_p$ (GeV) can be adequately
represented by the empirical relationship \cite{rs98,py00}

$$N(E_p,x)={S_0}{{E_p}\over {\epsilon}}e^{t_m}\left( {{t}\over {t_m}}\right)^{t_m}{e^{-t}}$$
where 
$$t_m(E_p)={{x_0}\over {\lambda}}\ln\left( {{E_p}\over{\epsilon}}\right) -1$$ 
where $S_0$=0.045, $\epsilon$=0.074, $x_0$= 37.2 $g~cm^{-2}$, ${t}={{x}\over
{\lambda}}$ and $\lambda$ =proton interaction mean free path in air (70
$g~cm^{-2}$). We chose the electron energy spectrum given by Zatsepin and
Chudakov which is independent of the stage of cascade development
\cite{rs98,zc62}.

$$F(E,x)\ dE = 0.75\ N(x){{d\delta}\over {1+\delta}}$$
where $F(E,x)$ and $N(x)$ are the number of electrons of energy $E$ and
all energies respectively at a depth $x$, $\delta$ = 2.3$E/\kappa$ and
$\kappa$ = 72 MeV with $E$ in MeV.  Even though this approximates well at
shower maximum in the energy range 20-300 $MeV$, the minor departure at
other depths affects only the shape of the lateral distribution of \v
Cerenkov photons and does not affect the present estimates significantly.
The integral electron energy spectrum is then derived:

$$F(E,x)=0.75\ N(x)\ \ln\left( {{1}+{{2.3}\over {\kappa}}E}\right)$$

Substituting $E_{th}$ for $E$, the number of electrons above the \v
Cerenkov threshold at a particular depth are calculated using the above
equation.

\begin{figure}
\centerline{\psfig{file=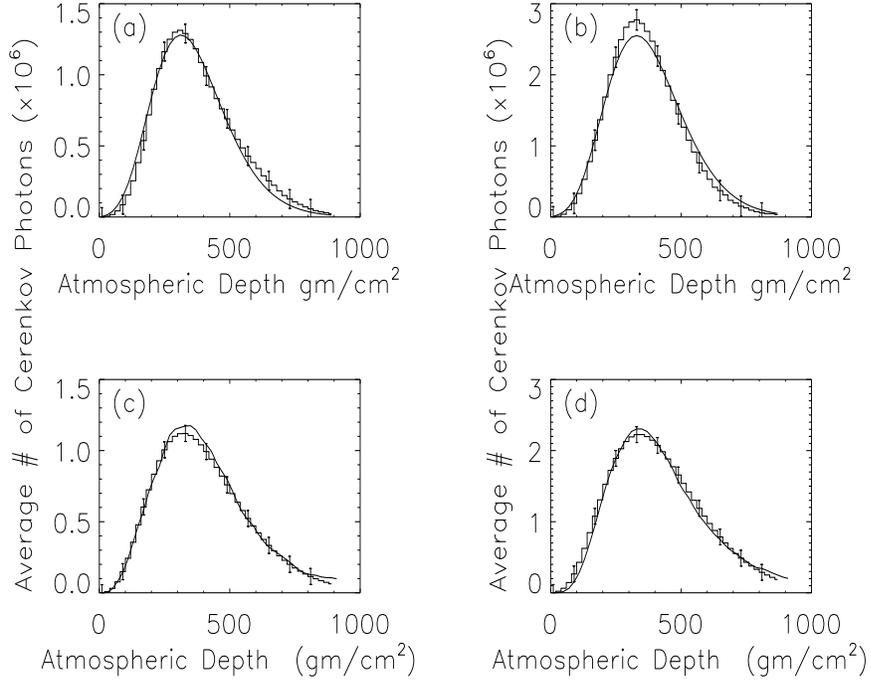,height=10cm,width=12cm
}}

\caption{\v Cerenkov photon growth curves in the atmosphere for $\gamma -$
ray of (a) 500 GeV \& (b) 1 TeV and protons of (c) 1 TeV \& (d) 2 TeV
primaries.  The histograms indicates the simulation results from CORSIKA
while the smooth curves indicates the results from analytical
calculations. The agreement between the two curves is good.
}
\label{fig1}
\end{figure}

The longitudinal \v Cerenkov photon profiles are then derived for $\gamma
-$rays of primary energy 50, 100, 250, 500 and 1000 GeV and protons of
energy 100, 200, 500, 1000 and 2000 GeV. The growth curves are simulated
using CORSIKA for two primary energies for $\gamma -$rays (500 GeV \& 1000
GeV) and protons (1000 \& 2000 GeV). Figure \ref{fig1} shows a comparison
of the simulated longitudinal profiles for $\gamma -$ rays (a \& c) as
well as protons (b \& d) with calculated profiles. The simulated profiles
(histograms) are averaged over 30 showers for $\gamma -$ rays and 50
showers for protons. The agreement between the simulated and calculated
longitudinal shower development profiles in terms of \v Cerenkov photons
is good for $\gamma -$ray as well as proton primaries of energies
considered here. Table \ref{tab1} summarizes the total number of \v
Cerenkov photons in these showers. The total number of \v Cerenkov photons
agree with that from simulations within 1 \& 2\% for $\gamma -$ray and proton
primaries respectively.

\begin{table}

\caption{Quantitative comparison of the simulated and calculated \v Cerenkov
photon growth curves in the atmosphere for $\gamma -$ray and proton primaries
at two primary energies.}

\label{tab1}
\vskip 0.3cm
\begin{tabular}{cccc}
\hline
Results  & Type of  & Energy of & Total \# of   \\
from  & primary  & primary   & \v Cerenkov  \\
       & & (GeV)     & photons ($\times 10^7$)    \\
\hline
 & $\gamma -$ rays &  500 & 2.25 \\
\cline{2-4}
   & Protons    &   1000   & 2.35 \\
\cline{2-4}
Simulations & $\gamma -$ rays & 1000 & 4.55     \\
\cline{2-4}
   & Protons    &   2000    & 4.81     \\
\hline
             & $\gamma -$ rays & 500  & 2.22\\
\cline{2-4}
   & protons    &  1000  & 2.35  \\
\cline{2-4}
Calculations & $\gamma -$ rays & 1000 & 4.56      \\
\cline{2-4}
   & protons    &  2000  & 4.72     \\

\hline
\end{tabular}
\end{table}

\begin{figure}
\centerline{\psfig{file=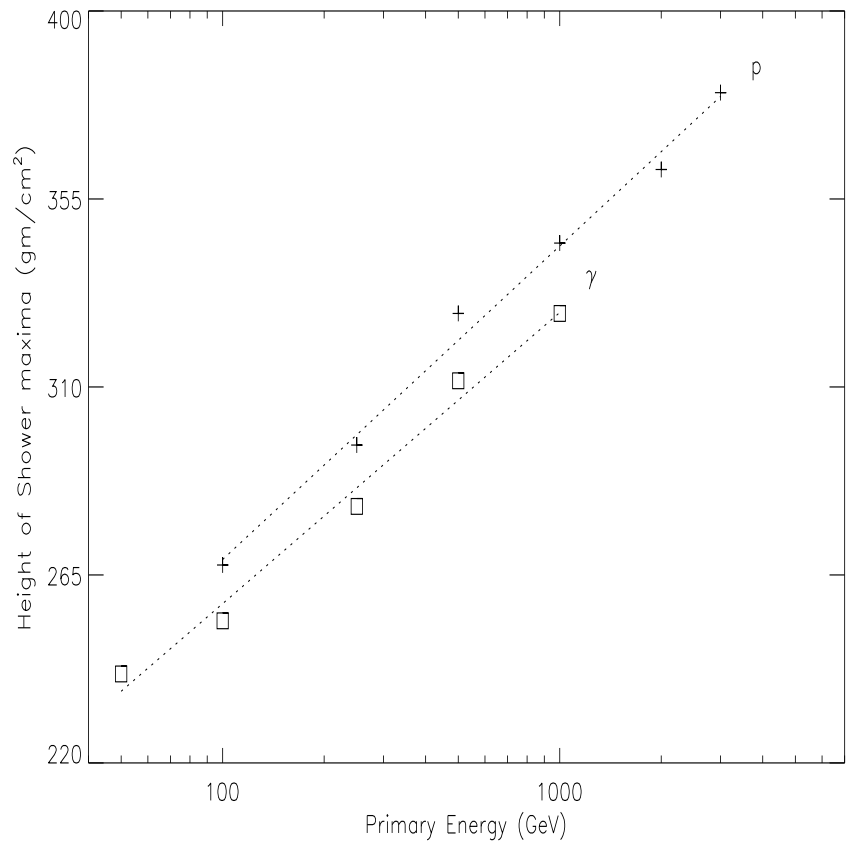,height=12cm,width=12cm
}}
\caption{ The variation of the depth of shower maximum with primary energy for
$\gamma -$rays and protons.
}
\label{fig5}
\end{figure}

The depth of shower maximum is expected to increase logarithmically with
primary energy both for electromagnetic as well as hadronic showers as
more cascade generations are required to degrade the secondary particle
energies \cite{rs98,py00}. Figure \ref{fig5} shows a plot of the depth of
shower maximum as a function of primary energy, both for $\gamma -$ray and
proton primaries in order to illustrate this point as well as to confirm
the correctness of the present calculations. The energy dependence of the
depth of shower maximum is parameterized as follows:

$$x_{max} = 30.3\ \ln E_{\gamma} +118.8\indent  g~cm^{-2}$$ for $\gamma -$rays
and,
$$x_{max} = 32.6\ \ln E_p +118.6\indent g~cm^{-2}$$
for protons.

\section{Photon attenuation in the atmosphere}

To study the attenuation of Cerenkov light in the atmosphere Elterman's
atmospheric attenuation model \cite{el68} is used, which provides the
attenuation coefficients for the Rayleigh and aerosol scattering as well
as ozone absorption in an altitude dependent form for the wavelength range
270-1260 $nm$ \cite{el68}.

Atmospheric attenuation model computes optical parameters spectrally and
with altitude as follows: (1) pure air attenuation parameters are
determined by utilizing Rayleigh scattering cross sections with molecular
number densities from standard atmosphere;  (2) ozone absorption
parameters are derived from coefficients applied to a representative
atmospheric ozone distribution; (3) seven sets of aerosol measurements are
compared and a profile of aerosol attenuation coefficients as a function
of altitude is developed. Tabulation permits calculations for vertical
path transmission at one kilometer intervals up to an altitude of 50 $km$,
individually for each attenuating component or for overall atmospheric
extinction (molecular + ozone + aerosol). Above 700 $nm$ light of night
sky (LONS) increases rapidly because of emission lines of OH and water
bands in upper atmosphere while the intensity of \v Cerenkov light drops
considerably( inversely proportional to the square of wavelength). At
lower wavelengths ($< 300~nm$) light undergoes strong absorption by ozone
molecules in the upper atmosphere.

The number of photons, integrated over the bandwidth of 300-550 $nm$
(dictated by the photo-tube band-width) and transmitted through a slab (1
$km$ thick) using the corresponding extinction coefficient is given by:

                                    $$ N=N_1 e^{-B}$$
where N is the number of photons transmitted through the slab with
absorption coefficient B and $N_1$ is the mean number of photons entering
the slab. For each altitude, the photon transmission spectrum is convolved
with the Cerenkov photon emission spectrum to get the transmitted photon
spectrum. Showers pass through several such slabs with different
transmission coefficients, resulting in a modified longitudinal
development profile for the Cerenkov photons that reach the observation
level.  The ratio of the total number of photons within the band-width in a
shower received at a particular observation level to the total photons
produced is defined as the transmission coefficient. The average
transmission coefficients ($T_c$) are listed in tables \ref{tab2} and
\ref{tab3} for $\gamma -$ray and proton primaries respectively for
Pachmarhi altitude. $T_c$ can be expressed as a power law in primary
energy, E as

\begin{equation}
T_c =a {E}^b
\end{equation}

\begin{figure}
\centerline{\psfig{file=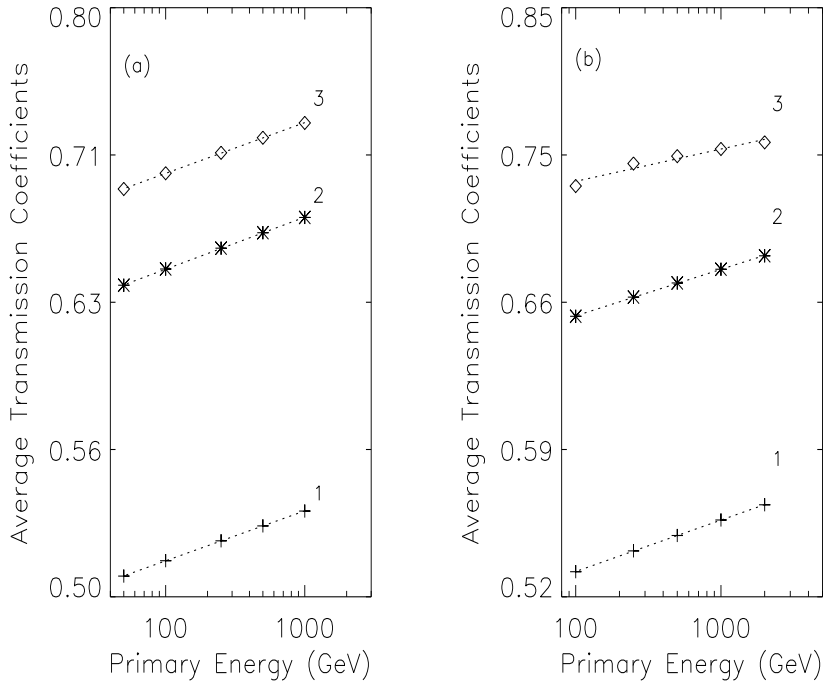,height=10cm,width=12cm
}}
\caption{Average transmission coefficient for \v Cerenkov photons in the
atmosphere for primary (a)  $\gamma -$rays and (b) protons of various energies.
The straight lines show the fits. The three plots in each panel correspond to
three observation levels (1) sea level, (2) Pachmarhi (1 $km$ a.s.l.) and
(3) 2 $km$ (a.s.l.) altitude.
}
\label{fig2}
\end{figure}

The values of $a$ and $b$ are listed in table \ref{tab4} both for $\gamma
-$ray and proton primaries at three different observation levels. The
variation in the average transmission coefficients for \v Cerenkov photons
in the atmosphere is shown in figure \ref{fig2} for (a) $\gamma -$ray and
(b) protons as a function of primary energy at three different observation
levels, illustrating the power law behavior.

\begin{table}

\caption{Shower size and average transmission coefficients corresponding
to Pachmarhi altitude for primary $\gamma -$rays of various energies
(band-width: 300-550 $nm$). }

\label{tab2}
\vskip 0.3cm
\begin{tabular}{ccccc}
\hline
\multicolumn{1}{c}{Primary} & \multicolumn{2}{c}{ \# of \v C }
&  \multicolumn{1}{c}{Average} & \multicolumn{1}{c}{Depth   of}\\
\multicolumn{1}{c}{energy} & \multicolumn{2}{c}{ Photons    }
&  \multicolumn{1}{c}{Transmission} & \multicolumn{1}{c}{shower max.}\\

(GeV)  &  produced  & transmitted & coefficient & ($g~cm^{-2}$)  \\
\hline
50  & $2.02\times 10^6$ & $1.29\times 10^6$  & 0.641 &241.3 \\
\hline
100 & $4.16\times 10^6$ & $2.70\times 10^6$  & 0.649 &254.0\\
\hline
250 & $1.08\times 10^7$ & $7.14\times 10^6$  & 0.660 &281.4   \\
\hline
500 & $2.22\times 10^7$ & $1.48\times 10^7$  & 0.669 &311.5  \\
\hline
1000& $4.54\times 10^7$ & $3.07\times 10^7$  & 0.677 &327.6 \\
\hline
\end{tabular}
\end{table}

\begin{table}
\caption{Shower size and average transmission coefficients corresponding to
Pachmarhi altitude for primary protons of various energies (band-width: 300-550
$nm$).
}
\label{tab3}
\vskip 0.3cm
\begin{tabular}{ccccc}
\hline
\multicolumn{1}{c}{Primary} & \multicolumn{2}{c}{ \# of \v C }
&  \multicolumn{1}{c}{Average} & \multicolumn{1}{c}{ Depth   of}\\
\multicolumn{1}{c}{energy} & \multicolumn{2}{c}{ photons    }
&  \multicolumn{1}{c}{transmission} & \multicolumn{1}{c}{shower max.}\\

(GeV)  &  produced  & transmitted & coefficient & ($g~cm^{-2}$)  \\
\hline
100 & $2.17\times 10^6$ & $1.42\times 10^6$  & 0.657 &267.4 \\
\hline
250 & $5.56\times 10^6$ & $3.70\times 10^6$  & 0.665 &296.1\\
\hline
500 & $1.15\times 10^7$ & $7.74\times 10^6$  & 0.673 &327.6 \\
\hline
1000& $2.37\times 10^7$ & $1.61\times 10^7$  & 0.681 &344.5\\
\hline
2000& $4.86\times 10^7$ & $3.34\times 10^7$  & 0.688 &362.1  \\
\hline
\end{tabular}
\end{table}

\begin{table}

\caption{Fitted coefficients $a$ and $b$ in equation 1 for $\gamma -$ray and
proton primaries at three different observation levels.
}

\label{tab4}
\vskip 0.3cm
\begin{tabular}{cccc}
\hline
Observation & Fitted & $\gamma -$rays & Protons  \\
level $km$ a.s.l.& coefficients &        &            \\
\hline
 0   & $a$ &   0.475  &  0.487      \\
(Sea level)  & $b$ &   0.018  &  0.019      \\
\hline
1           & $a$ &   0.598  &  0.608      \\
            & $b$ &   0.018  &  0.017      \\
\hline
2           & $a$ &   0.646  &  0.697      \\
            & $b$ &   0.018  &  0.012      \\
\hline
\end{tabular}
\end{table}

The increase in $T_c$ with primary energy implies that higher energy
primaries penetrate deeper in the atmosphere and hence pass through lesser
air mass. The proton primaries reach the shower maximum lower down in the
atmosphere compared to a $\gamma -$ray primary of the same energy since
the interaction mean free path in air for the former is nearly twice the
radiation length. As a result the protons have marginally larger ($\sim
1.2\%$ at 100 GeV to $\sim 0.6\%$ at 1 TeV) average transmission
coefficient compared to a $\gamma -$ray primary of the same energy.

\begin{figure}
\centerline{\psfig{file=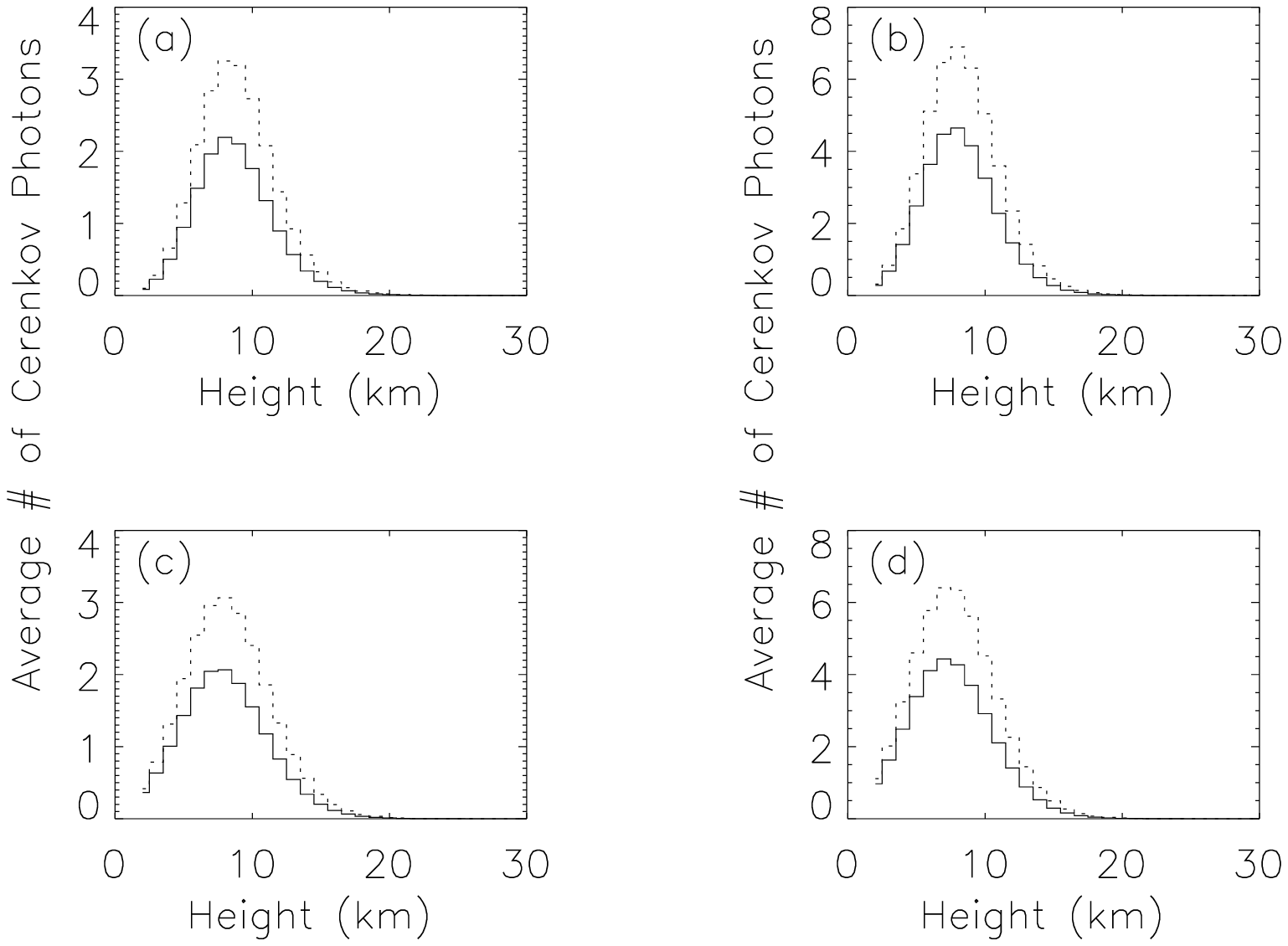,height=12cm,width=12cm
}}
\caption{A comparison of the longitudinal development profiles of \v Cerenkov
photons at Pachmarhi level with and without atmospheric attenuation correction.
$a$ \& $b$ are for $\gamma -$ray primaries of energy 50 GeV and 1 TeV
respectively while $c$ \& $d$ are for proton primaries of energy 100 GeV and
2 TeV respectively (band-width: 300-550 $nm$).
}
\label{fig3}
\end{figure}

The attenuation of optical photons due to Rayleigh and aerosol scattering
is more significant at lower altitudes. As a result the \v Cerenkov photon
transmission is expected to depend on the observation level as also seen
in figure \ref{fig2}. The average photon transmission coefficient increases
almost linearly with decreasing atmospheric depth and the rate is
$\sim 0.1\%$ for every 100 $g~cm^{-2}$ of air.

Figure \ref{fig3} shows the longitudinal development profiles of \v
Cerenkov photons at production in the atmosphere and that for the photons
detected at the observation level ($i.e.$ ``without'' and ``with''
atmospheric attenuation correction) for the same wavelength band. The 
differences in the two profiles is due to the wavelength dependent absorption
of \v Cerenkov photons in the atmosphere. From the plots it is clear that the
shape of the longitudinal profile remains largely unchanged (except for the
total number of photons). The range of atmospheric heights from which the
detected photons originate also remains unchanged as well, irrespective of
the primary energy or species. Thus the height of shower maximum remains
unchanged due to atmospheric attenuation.

\begin{figure}
\centerline{\psfig{file=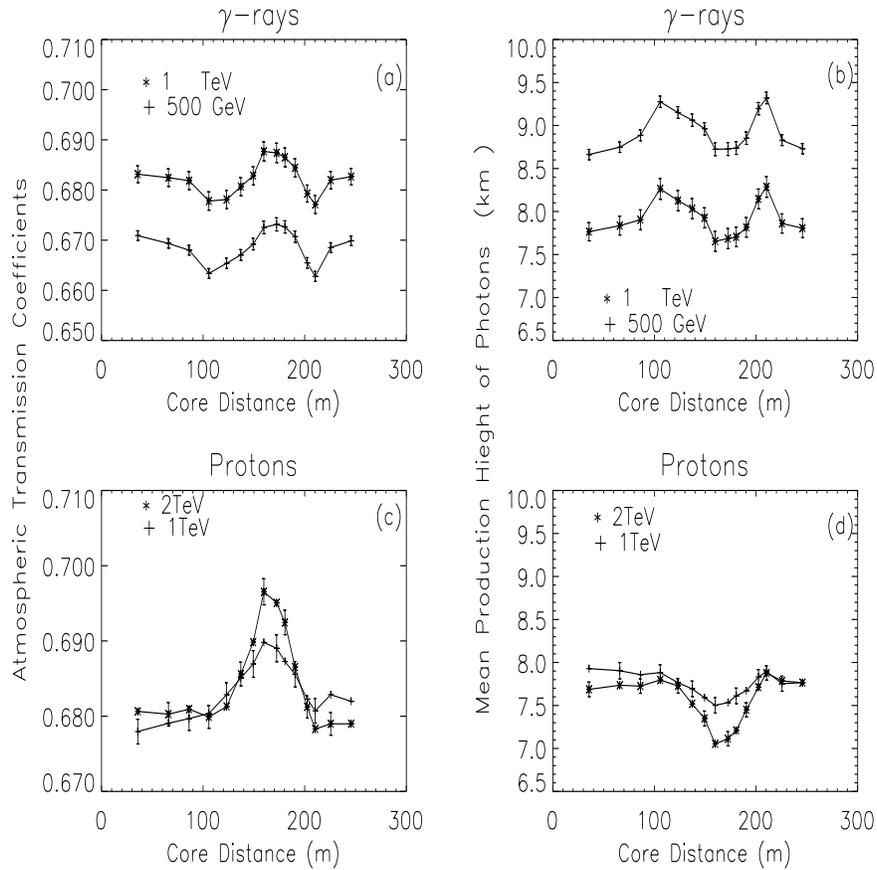,height=12cm,width=12cm
}}

\caption{Radial dependence of the \v Cerenkov photon transmission
coefficients in the atmosphere for (a) $\gamma -$ray and (c) proton
primaries of two different primary energies as shown. Also shown are the
radial dependence of average production heights for each of the primaries
(b \& d respectively) and their energies mentioned above.
Simulation results
are averages over 43 \& 17 $\gamma -$ray showers of primary energy 500 GeV and
1 TeV respectively while 47 showers are used for 1 \& 2 TeV proton primaries.
}

\label{fig4}
\end{figure} 

The average path-length of \v Cerenkov photons reaching different core
distances differ resulting in different attenuation. Figure \ref{fig4}
shows such a radial variation of the transmission coefficient for \v
Cerenkov photons generated by $\gamma -$rays and protons of two different
energies. Also shown are the radial variation of mean production heights.
It can be seen that the radial variation of transmission coefficients is
rather small but varies inversely as the mean production height.

\section{\v Cerenkov photon spectrum}

\begin{figure}
\centerline{\psfig{file=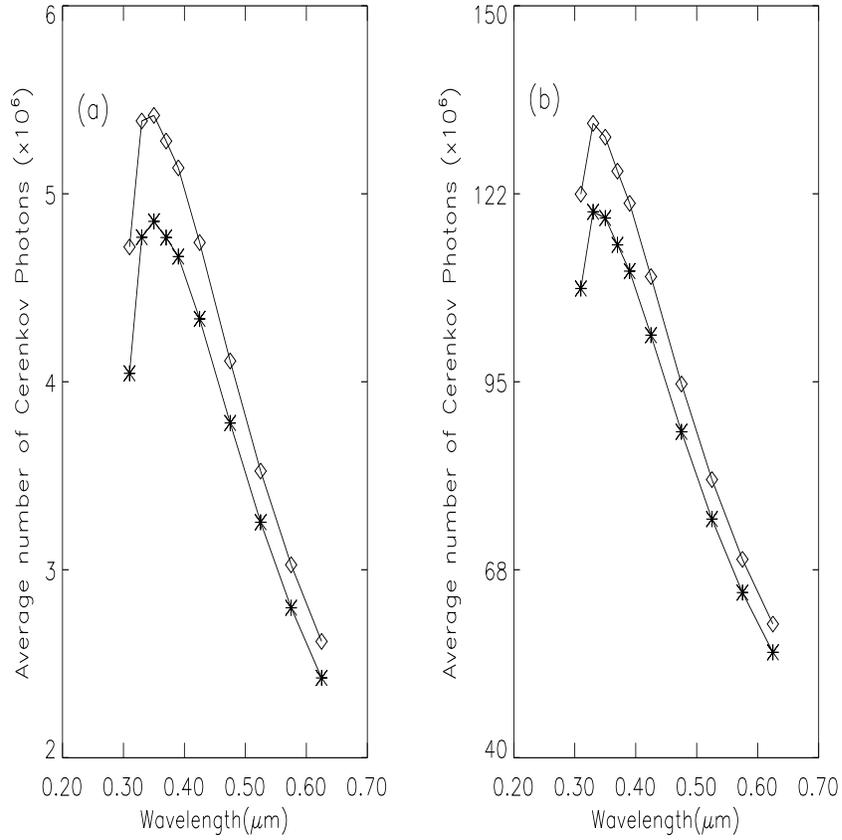,height=12cm,width=12cm
}}

\caption{Typical \v Cerenkov photon spectra at Pachmarhi level. (a) shows
the spectra from $\gamma -$ray primaries of energy 50 GeV and 100 GeV
protons while (b) shows those for 1 TeV $\gamma -$ray \& 2 TeV proton
primaries. The symbols used are * for $\gamma -$rays and diamond for
protons.}

\label{fig6}
\end{figure}

The \v Cerenkov photon spectrum at the observation level is an important
input for the design of an atmospheric \v Cerenkov experiment. The
fraction of the photon spectrum bracketed by the photo-multiplier
bandwidth has an important bearing on the sensitivity of a TeV $\gamma
-$ray telescope. Hence we computed the photon spectrum at different
observation levels for $\gamma -$ray and proton primaries of various
energies. The bandwidth considered in our calculation is 270-550 $nm$.
Figure \ref{fig6} shows photon spectra, corresponding to Pachmarhi level,
generated by (a) 50 GeV $\gamma -$rays \& 100 GeV protons (* \& diamond
respectively) as well as (b) 1 TeV $\gamma -$rays \& 2 TeV protons. It can
be seen from the figure that the wavelength at peak intensity is a
function of the primary energy in both the cases. It shifts from around
350 $nm$ at 50 GeV to around 330 $nm$ at 1 TeV. This is because the higher
energy primaries reach the shower maximum lower down in the atmosphere
compared to lower energy primaries. As a result, absorption of shorter
wavelength photons which is primarily due to atmospheric Ozone is
comparatively larger for lower energy primaries.

\subsection{Fraction of UV component at various altitudes}

The UV filters were used in atmospheric \v Cerenkov experiments
\cite{go88} whose purpose was two-fold: firstly, it helps reduce night sky
background while minimizing the loss of \v Cerenkov light in the blue and
near UV range. This would provide better stability at higher photo-tube
gains which in turn will improve the sensitivity of an atmospheric \v
Cerenkov telescope. Secondly, it could serve to identify showers with a
larger UV fraction in the \v Cerenkov light. Since the \v Cerenkov light
generated by proton primaries traverse lesser air mass as compared to
$\gamma -$ray primaries it is expected to have a larger UV content. This
property could be exploited to discriminate against hadronic showers
\cite{zy81}.

Hence the UV fraction in \v Cerenkov light at observation level is a
useful signature. So we estimated this fraction for primaries of various
energies. We divided the \v Cerenkov photon spectrum as seen at the
observation level into two groups $viz.$, the UV range comprising
wavelengths 270-300 $nm$ and the visible range comprising the wavelength
band 300-550 $nm$. The ratio of the number of photons in the UV range to
that in the visible range is defined as $R_p$ for proton primaries and
$R_\gamma$ for $\gamma -$ray primaries. Table \ref{tab5} summarizes the
$R_\gamma$ and $R_p$ values for different primary energies as seen at
three different observation altitudes.

\begin{table}

\caption{Ratio of UV (270-300 $nm$) to visible (300-550 $nm$) component in the
\v Cerenkov light detected at three observation altitudes as a function of
$\gamma -$ray and proton energies.
}

\label{tab5}
\vskip 0.3cm
\begin{tabular}{ccccccc}

\hline
\multicolumn{1}{c}{Primary} & \multicolumn{2}{c}{ Sea level  }
&  \multicolumn{2}{c}{1 $km$ } & \multicolumn{2}{c}{2 $km$}\\

energy (GeV)&$R_\gamma$  & $R_p$   & $R_\gamma$  & $R_p$   &$R_\gamma$&$R_p$ \\
$E_\gamma /E_p$& (\%)  &   (\%)      &   (\%)        &  (\%)   &  (\%)  & (\%)\\
\hline
50 /100& 8.23        & 9.57        & 8.27          & 9.67    & 8.29   & 9.77  \\
\hline
100/200& 8.26        & 9.59        & 8.29          & 9.69    & 8.33   & 9.80  \\
\hline
250/500& 8.29        & 9.60        & 8.32          & 9.71    & 8.37   & 9.85  \\
\hline
500/1000& 8.31       & 9.61        & 8.34          & 9.72    & 8.41   & 9.88  \\
\hline
1000/2000& 8.32      & 9.61        & 8.36          & 9.73    & 8.44   & 9.91  \\
\hline
\end{tabular}
\end{table}

\begin{figure}
\centerline{\psfig{file=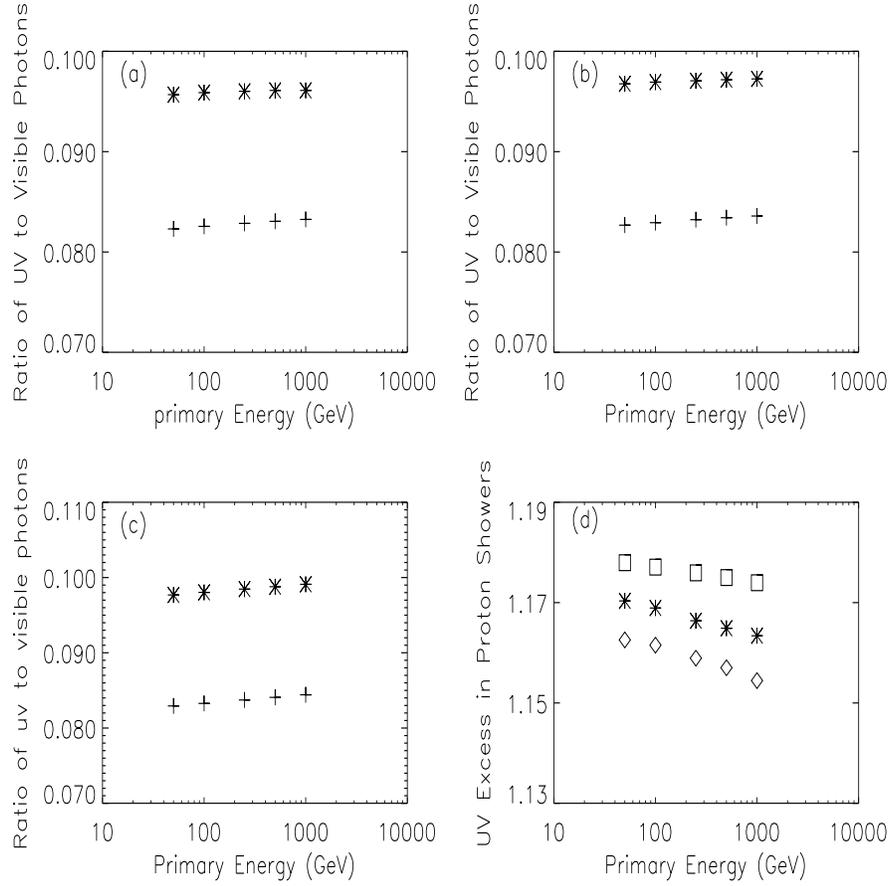,height=12cm,width=12cm
}}

\caption{Variation of the ratio of UV to visible fraction in \v Cerenkov light 
generated by protons (*) and $\gamma -$rays (+) as a function of primary
energy, at three observation levels: (a) sea level (b) Pachmarhi and (c) 2
$km$ a.s.l. (d) The relative excess in the UV content in hadronic primaries as
a function of primary energy expected at three different observation
levels, in the same order from bottom to top. }

\label{fig7}
\end{figure}

Figure \ref{fig7} shows a plot of $R_p$ and $R_\gamma$ as a function of
primary energy for three different observation levels. Also shown in
figure \ref{fig7} are the relative excess of UV content in proton showers
\emph{vis-a-vis} $\gamma -$ray showers, as a function of primary energy at
three different altitudes.

It can be readily seen from the figure that sensitivity to UV excess is
better at higher altitudes as compared to sea level.

\begin{figure}
\centerline{\psfig{file=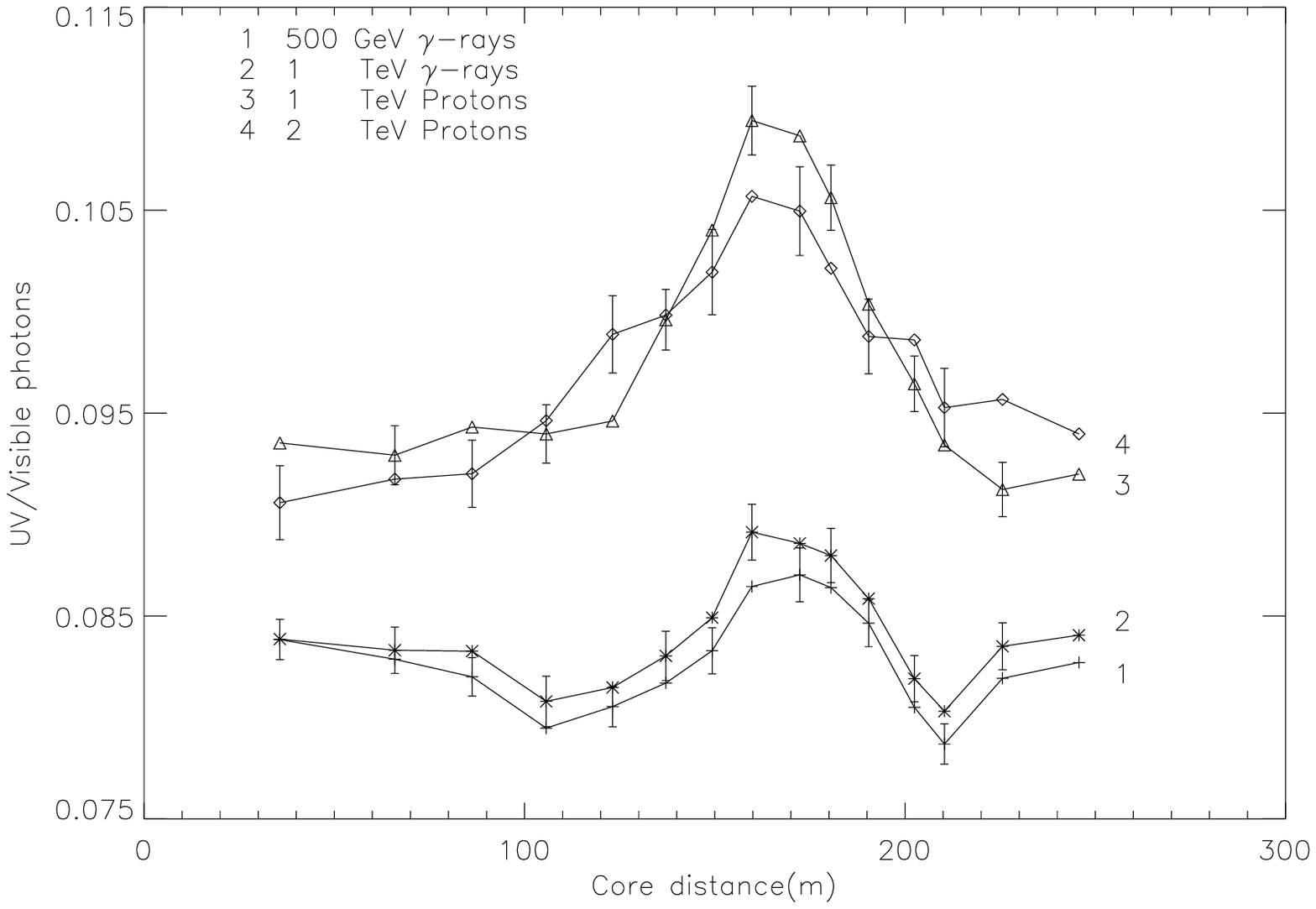,height=12cm,width=12cm
}}

\caption{Radial variation of the ratio of UV to visible photons in \v Cerenkov
light generated by protons (1 \& 2) and $\gamma -$rays (3 \& 4) at two primary
energies. Alternate error bars are shown for clarity.}

\label{fig8}
\end{figure}

Figure \ref{fig8} shows radial variation of the ratio of UV to visible
photons for $\gamma -$ray and proton primaries at an altitude of 1
\emph{km}. The relative UV fraction is marginally higher around the hump
region since the photons in this region are contributed by higher energy
electrons which have shortest path-length in air. The radial dependence of
the UV fraction is more pronounced for proton primaries for reasons already
mentioned before.

\section{Discussion and Conclusions}

In spite of the complexity of the interaction kinematics of high energy
cosmic rays in the atmosphere, the average behavior as derived from
detailed simulation studies agrees reasonably well with analytical
calculations. This demonstrates that the simulation package does
take into account almost all the interaction characteristics giving
credence to the conclusions drawn from simulation studies. The position of
shower maximum obtained by us agrees well with those of Miller \&
Westerhoff \cite{mw98}.

Armed with the above result we proceed to make analytical calculations for
the production of \v Cerenkov light produced by both $\gamma -$rays and
protons of various energies at three different observation altitudes. This
is ideal to study the average shower properties as this is much faster
than the detailed simulations. We then apply a detailed wavelength
dependent corrections due to the attenuation of \v Cerenkov photons during
their propagation in the atmosphere. This has been done for $\gamma -$ray
and proton primaries of various energies and for three observation levels.
It has been found that the fraction of the transmitted \v Cerenkov photons
at an observation level of 2 $km$ above sea level is about 36\% more than
that at sea level. This is primarily because the average distance to
shower maximum is reduced resulting in lesser absorption. Even though the
effective collection area is reduced at a higher altitude, the light
intensities are larger near the shower core. In addition, increased UV
content in the \v Cerenkov light makes higher observation levels better
suited for atmospheric \v Cerenkov work. Simulations carried out for Mt.
Hopkins observatory site (altitude of 2.3 $km$) suggest an advantage
factor in the range 1.2-1.4 compared to sea level \cite{ph89}, is
consistent with the present result. We also derived the observed \v
Cerenkov photon spectrum for a $\gamma -$ray of energy 3 TeV at an
observation altitude of 2 $km$. This spectrum agrees well with that
reported by Mirzoyan, {\it et~al.}\cite{rm94} for HEGRA observation level.

The use of a UV filter to preferentially accept \v Cerenkov light has been
suggested even though its efficiency in improving the signal to noise ratio
of an atmospheric \v Cerenkov telescope  is in doubt \cite{ph89}. On the
other hand pushing the sensitivity to near UV wavelengths does improve the
system sensitivity \cite{go88,pe93}. Also using suitable filters to
suppress the photo-tube sensitivity in the visual band, existing imaging
telescopes have been modified so that observations could be made during
moderately moonlit nights. This technique has been shown to be successful
in detecting $TeV ~\gamma -$ray signal from Crab and Mkn 421 despite
higher energy threshold and reduced sensitivity \cite{sa96,ch97}.

It has been observed by Zyskin,{\em et al., }\cite{zy81} that the
relative signal strengths in the UV range (200-300 $nm$) to that in the
visible range (300-600 $nm$) increases with the angular size of the image
in the visible. It increases up to $(8\pm 1.8)\%$ for large
images($\sqrt{a.b}\sim 0.5^\circ $ where $a$ and $b$ are the semi-major
and semi-minor axis of the image in the optical). This could be due to the
fact that those showers whose maximum is closer to the observation level
produce larger \v Cerenkov images.  This is equivalent to raising
observation level which effectively reduces the attenuation of UV photons
as discussed in \S4.1 From the present calculations the relative signal
strength at a primary proton energy of 4 TeV (energy threshold of Zyskin
{\em et al., }\cite{zy81}) is estimated to be $\sim 10\%$ which is
quite consistent with their measurement.  It may be noticed that the
wavelength ranges used by Zyskin {\em et al., } is broader in UV and
while the visible band in our estimate was modified to match theirs.
However the observed \v Cerenkov photon spectrum falls steeply below $\sim
300~nm$, as could be seen in figure \ref{fig6}. Extending the wavelength
range below 270 $nm$ does not change the UV content significantly.

We find that the relative strengths of UV photons to the visible is higher
in the case of proton primaries by about 16\% at 50 GeV and decreases to
12\% at 1 TeV. Hence, the hadron discrimination efficiency based on the UV
content in \v Cerenkov light is relatively better at lower primary
energies and at near core distances.  Thus, measurement of relative UV content
of a shower could be a good parameter in order to discriminate against
hadrons especially for large ground based arrays with low energy
thresholds ($\sim 20-50~GeV$) and employing the wavefront sampling
technique \cite{on96,cel96,arq97,oc97,tum90}. However one has to keep in
mind that for this technique to be successful one has to make a very
accurate (better than $\sim 1\%$) estimate of the UV and visible light contents 
of the shower.

\end{document}